# POWER CONTROL IN REACTIVE ROUTING PROTOCOL FOR MOBILE AD HOC NETWORK

Maher HENI[1] and Ridha BOUALLEGUE[2]

[1,2]Innovation of COMmunicant and COoperative Mobiles Laboratory, INNOV'COM
Sup'COM, Higher School of Communication, Ariana, Tunisia
[1] henimaher@gmail.com
[2] ridha.bouallegue@gmail.com

## ABSTRACT

*The aim of this work is to change the routing strategy of AODV protocol (Ad hoc On Demand Vector) in order to improve the energy consumption in mobile ad hoc networks (MANET). The purpose is to minimize the regular period of HELLO messages generated by the AODV protocol used for the research, development and maintenance of routes. This information is useful to have an idea about battery power levels of different network hosts. After storing this information, the node elect the shortest path following the classical model used this information to elect safest path (make a compromise) in terms of energy. Transmitter node does not select another node as its battery will be exhausted soon.*

*Any node of the network can have the same information's about the neighborhoods as well as other information about the energy level of the different terminal to avoid routing using a link that will be lost due to an exhausted battery of a node in this link.*

*Analytical study and simulations by Jist/SWANS have been conducted to note that no divergence relatively to the classical AODV, a node can have this type of information that improves the energy efficiency in ad hoc networks.*

## 1. INTRODUCTION

A Mobile Ad-hoc Network (MANETs) [1] is a collection of autonomous nodes or terminals that communicate together by forming a multi-hop radio network and maintaining connectivity decentralized. The nodes can move and their network topology may be temporal. Each node acts as a customer, server and router. In such network, there is no centralized administration. Each node can join the network or it leave at any time.

Routing protocols in such networks can be classified mainly into three categories:

• Proactive routing protocols: They are based on the same principle as wired networks routing. Paths in this type of routing are calculated in advance. Each node maintains multiple routing tables by exchanging control packets between neighbors. Indeed, if a node wants to communicate with one another, it has the ability to view local routing table and create path it needs. OLSR [2](Optimized Link State Routing) and FSR [3](Fisheye State Routing) are examples of proactive routing protocols.

• Reactive routing protocols: On the Contrary of proactive protocols, reactive protocols calculate the route on request. If a source node needs to send a message to a destination node, then it sends a request to all members of the network. After receiving the request, the destination node sends a response back to the source. However, the routing application generates a slow pace because of the research paths which can degrade application performance. Such protocol has the disadvantage of being very costly in terms of energy and packets transmission when determining routes but has the advantage of not having to hold unused information in routing tables. AODV [4] is an example of reactive protocols which are described below.

• Hybrid routing protocols: Hybrid routing protocols or "mixed" combine the previous two types of routing (proactive and reactive). The proactive protocol is applied in a small area around the source (limited number of neighbors), while the reactive protocol is applied beyond this perimeter (distant neighbors). This combination is performed in order to exploit the advantages of each method and overcome their limitations. ZRP [8] (Zone Routing Protocol) and CBRP [7] (Cluster Based Routing Protocol) are two major examples of hybrid protocols.

One of the major and most critical factors in ad hoc networks is the limited battery energy. A large amount of works is focused on this setting to reduce the consumption of batteries. The waste of energy may be due to the regular exchange of unnecessary control messages to have more reliability.

AODV (Ad hoc On-Demand Distance Vector Routing Protocol) is a reactive routing protocol designed by Charles E. Perkins and Elizabeth M. Royer [4]. This protocol uses four types of control messages in the aim to send data packets. The first type is HELLO messages. This type of messages, exchanged periodically to maintain a neighborhood base. RREQ, RREP, RRER, are used to establish a path to destination when any node wants to send a data. This number of control packet has a signified effect of the waste of resources.

To overcome the problem of energy consumption in this protocol, we designed a new solution that reduces the HELLO messages number exchanged and to include the factor of energy consumption that will be useful later for the routing messages. Firstly, we minimize the exchange number of Hello messages. Secondly, we replace the regular periodic instant of sending hello message by another proportionally to the energy stored in the battery of the node. The node receiver of this hello message, do the inversely action to extract information proportionally to the node sender energy, and the same information enclosed in hello message.

Insert this parameter does not affect the operation or the information included in messages exchanged and then we can obtain new information that we can use to elect path. We call the new protocol PC-AODV (Power Control AODV).

This paper is organized as follows: we present the AODV routing protocol in Section 2. In Section 3, we detail the related work, we expose the used model and parameters in section 4, and we formalize our solution and we present the new protocol called PC-AODV in section 5. In the rest of the paper, we illustrate, in section 6, an analytical comparative study between the classical and the new protocol. In section 7, we present a simulation evaluation, of the tow protocol using JiST/SWANS simulator. We conclude this paper and present future work in Section 8.

## 2. AD HOC ON-DEMAND DISTANCE VECTOR ROUTING PROTOCOL

### 2.1. Overview

AODV [4] [5] is a reactive protocol that is based on the concept of distance vector routing protocols as its name mean. The algorithm of AODV is inspired from the combination of a proactive and a reactive protocol [24, 25]. The path discovers and maintains is similar to the process used in DSR [26]. The uses of HELLO message exchange to establish a neighborhood base and sequence number method are used in DSDV [27]. AODV present more performance in static and bulky networks. These factors present major challenges to MANETs routing protocol researchers.

AODV performs route discovery request and saves only used routes in the routing table. It use four different control message called HELLO, RREQ, RREP, and RRER message. In order to transmit data packets, it broadcasts a route request RREQ (Route REQuest message) in the wholly networks. Three cases are possible upon receipt of a RREQ message by any node. In the first, if the node that received this message provides a route to the requested destination in its routing table, it responds with another type of message RREP (Route REPly message). In the second case, if it hasn't information about the destination, it will retransmit the message to its neighbors that have not yet received. If all the neighbors have received the same message and/or the node has lost the connection, it responds with an error message RERR (Rout ERRor message). After receiving a reply message, the source node starts sending data packets along the shortest path.

Other than these messages, AODV uses only one type of periodic message is HELLO message, in order to maintain the Neighborhood basis.

In either case, the source node waits for a predefined timeout, the route establishment response to the destination, and then it retransmits another RREQ by increasing the maximum number of hops (TTL: Time To Live). If after repeating this process a limited number and the source get nothing, it declares the absence of this destination.

To maintain routes, AODV use an ACTIVE_ROUTE_TIMEOUT (ART) that equal to 3 second [28]. If and defined routes between tow nodes, is not used within this period, then this node is not sure if this route is yet available or not, it rebroadcast a RREQ if needs

## 2.2. Motivation

To exchange these types of route establishment messages, each node periodically exchange HELLO messages to maintain a neighborhood base and the routing table. Since the regular exchange of both control messages amplifies considerably the energy consumption, and bandwidth.

Generally, MANETs are characterized by limited energy and bandwidth. With the exchange of this considerable number of control messages to establishment of routes, this aggravates the resources and performances, precisely in the case of bulky networks. One of major causes of exchange of these message is the lost of paths, result of the exhausting of a node battery. To overcome this problem, we suppose that all nodes composed the networks have information about the energy stored in the batteries of its neighborhood. It avoid the routing using a node that it battery will be exhausted and take into account the nodes that can be used.

To do this, we propose a mechanism that reduces the exchange of this type of message and use it to inform about the energy state of the other nodes in keeping the same performances of the standard protocol. Using hello message, all nodes exchange a new type of information about the energy stored in battery, without changing the fields composed this message, and simply tuning the instant of sending it.

## 3. RELATED WORK

The work done in this context could be grouped into two major groups; the first describes methods for reducing energy consumption in the AODV protocol with diversifying the routing strategy, and the second present's methods to reduce numbers of control messages in order to reduce the cost of consumption of energy.

In [14] authors propose a new version of AODV called (MAODV) derived from the AODV routing protocol by considering the bit error rate (BER) at the end of a multi-hop path as the metric to be minimized for route selection. In [15], authors integrated the transmit power control and load balancing approach as a mechanism to improve the performance of on-demand routing with energy efficiency. M.Veerayya, V. Sharma and A. Karandikar propose in [16] a cross-layering approach to exchange information about the residual energy in nodes to perform quality of service. In [17] a new mechanism is proposed to set a timeout for a path. A path considered broken if a node leave by following the exhaustion of its energy. In [18] authors integrate the runtime battery capacity in routing protocol and the estimated real propagation power loss, obtained from sensing the received signal power. This solution is independent of location information and using the propagation, they estimate the energy loosed. Another type of the proposed work which aims to reduce the overhead of AODV to achieve energy efficiency, as described in [19]. Authors propose a new method in order to reduce overhead in AODV in urban area by predicting links availability. By predicting neighbor nodes positions it can be determined probability of link failure.

In [20] S.B. Kawish, B. Aslam, S. A. Khan studies the behavior of AODV in a fixed networks and those exhibiting low mobility with a view to highlight the reasons for reducing overhead and then reduce the energy consumption. The same authors present in [21] an improvement in their idea of using route timeout adjusted to reduce the overhead.

In [22] Authors propose a new version of AODV an on-demand routing algorithm based on cross-layer power control termed as called CPC-AODV (Cross-layer Power Control Ad hoc On-demand Distance Vector) taking account of the geographic location of nodes, the energy of packet transmission. Furthermore, the approach presented in [23] consists of an algorithm that enables packet forwarding misbehavior and Loss Reduction based detection through the principle of conservation of flow on the routing protocol group nodes.

First, unlikable the other proposed solution, our protocols, does not minimize the number of messages or the overhead, or use geographic coordinates of the nodes or the channel access using the MAC layer. Our solution simply changes the periodicity by random time for the receiver and set by the power level of the node battery the transmitter. This is an important feature and has a profound effect on energy consumption which could sustain the behavior of protocol. It is an available approach to incorporate routing protocols with power control in ad hoc networks.

## 4. USED MODEL

We use a network composed by four nodes (node A, B, C and D) with bidirectional or symmetric links between them. The communication range is circular with a diameter of 250 meters.

Table 1.  Used Variables.

| Variable | Designation |
|----------|-------------|
| $E_x$ | Energy stored in the node x battery |
| $E_R$ | Resultant energy |
| $K_x$ | 1/Ex |
| HELLO(x) | Message HELLO |
| $H_{ACK}(x)$ | Hello message acknowledgment |
| $T_{ACK}$ | reception time of HELLO message acknowledgments |
| Δt | acknowledgment period |
| HI | HELLO_INTERVAL |
| Nn | Node's neighborhood number nodes |

Our goal is primarily to have an idea about the quantity of energy stored in batteries for neighbors. The parameters used to define the model are defined in Table 1. The topology used is shown in Figure 1.

In our model, we decrease the number of HELLO messages by increasing the time between two messages. Assuming that the period between two successive HELLO messages is proportional to the neighbors number according to the equation (1).

$$HI_{PC-AODV} = N_n * HI_{AODV} \qquad (1)$$

We will reduce this interval to receive Acknowledgment that have the same content as the hello messages, but which allows to know the battery level of the other nodes.

Ki is inversely proportional to the battery's energy. Assuming that the máximum level of the battery power is 15Kw [29], in our example Kc is the node C factor,  then the level of it battry power  is

$$E_c = \frac{1}{K_c} * E_{Max} = \frac{1}{6} * 15Kw = 2,5Kw$$

Respectively

$$E_B = 5Kw, E_A = 7,5Kw, E_D = 3,75Kw$$

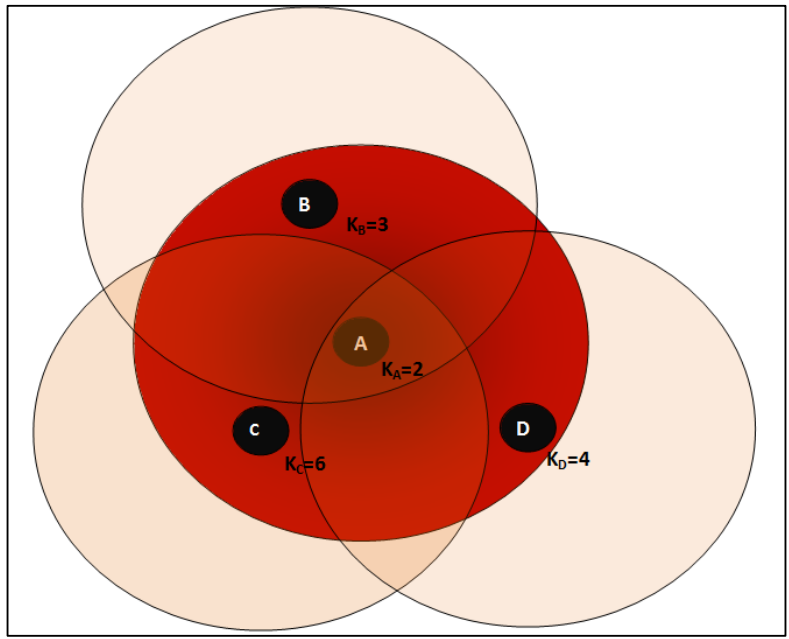

Figure 1. The neighbourhood topology used for the description

After receiving a hello message from node A, all node start send a Hello acknowledgments at an instant proportionally to the factor K. then, Node B will send, the first, an acknowledgments to A, node D and then node C (see Figure 2). Using this concept, node A do the inversely process to extract the level of neighbour's battery.

## 5. PC-AODV CONCEPT

Our solution is illustrated in Figure 2. After sending a Hello message, the node A starts receiving acknowledgments from its neighbors. The parameter δt is assumed known by all nodes and is defined in the HELLO message. We chose a parameter K the inverse of the energy stored in order to have the first acknowledgment of the node that has the maximum energy, and either receive or not the final acknowledgment at the end of the period. If the energy of a node is negligible, for example, you will not receive acknowledgment during the period $T_{ACK}$.

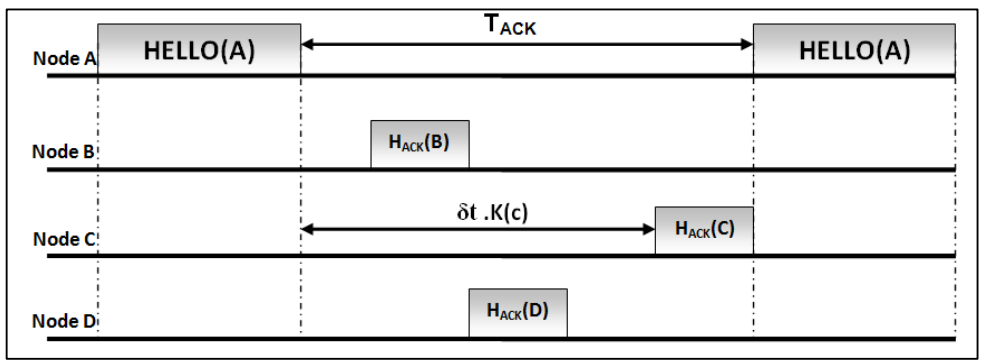

Figure 2. HELLO transmission and reception model

To better understand the phenomenon, we take the example of figure 1where K (B)=3, K(C) = 6 and K(D) = 4.where

$$K(i) = \frac{1}{node\_i\_Battery's\_level}$$

We assume that node A select δt = 2.

Thus the sent start time to of these nodes are:

$t_C$ = δt x K(C) =  2 x 6 = 12 Time_unity.

$t_B$ = δt x K(B) =  2 x 3 = 6 Time_unity.

$t_C$ = δt x K(D) =  2 x 4 = 8 Time_unity.

So the first node who starts sending is B , the next one is D , then node C.

After receiving various Hello acknowledgments node A registers and updates the localization information and battery energy in the neighborhood base. An Acknowledgement of HELLO message sent by a neighbor B and is itself another HELLO message generated by this node and containing the same information as classical message. The only difference is the non-periodicity of this message in order to have the desired information. Information about energy can be further used for the dissemination of other control messages or data. In the event that a node in the neighborhood has a battery exhausted, the node A avoids the flow of messages using this node.

After collecting information concerning the batteries level, a node source then chooses more than the shortest path, the safest path. Indeed, after selecting of shorter path,and sending a data message, a node can be left this way and so the link considered broken and then we have loss of message. However, in the case of safe path the source nodes can elect the path that contain no node with an exhausted battery.

## 6. ANALYTICAL COMPARATIVE STUDY

In this section, we present a comparison between AODV and PC-AODV. We compare analytically the two concepts in the same condition and parameters.

To prove the energy efficiency protocol for PC-AODV, we consider, as a metric, the energy required for the correct transmission of a packet from mobile node i to node j. On the first hand we take the energy Eij formula presented in [9] and [10].

$$E_{i,j} = \frac{MP_i}{RP_c(\delta_{i,j})}$$

(2)

Where M denotes the length of the packet, Pi is the transmission power of node i, R represents the data transmission rate and Pc(δij) is the probability of correct reception of a packet from node i to node j, with δij equal to the SIR ( Signal to Interference Ratio) [30] of link (i,j). This expression depends on the transmission quality and characteristics of link between the two nodes.

On the other hand, we take the second formula developed in [11]

$$E(p,n_i) = P(p,n_i) * t_p = i * v * t_p \qquad (3)$$

Where P(p,ni) is the power to transmit a packet $p$ by any node ni, v is the node battery voltage and, $i$ is the current (in Ampere).

i *v is the constant party of the equation and dependent on the technical characteristic of the host. Using [12], the time taken to transmit the packet p (in sec) is presented by (2):

$$t_p = \frac{p_h}{6*10^6} + \frac{p_d}{54*10^6} \qquad (4)$$

ph is the number of bits of the packet header, and pd is the payload presented in [17] by the following formula:

$$p_d = 256(228(data) + 8(UDP) + 20(IP)) \qquad (5)$$

In order to compare the two concepts, we consider two nodes x and y and assume that the parameters for message transmission are the same in both cases. Indeed, we take both equations (2) and (3) representing the energy, if we assume that the conditions are the same (same battery level, even message length in equation (3), the same transmission condition, even, message length for the equation (2)). Thus, we consider that the energy transmission of a HELLO or an acknowledgment message from source x to destination y in our solution equal to an accurate value C (equation (6)), since both have the same length and sent in the same condition.

$$E_{x,y} = C \qquad (6)$$

According to [9] and [10] the resulting energy ER (Resultant Energy) required for the routing of messages over a period t is determined by

$$E_R = \sum_{t,(\forall (i,j) \in N)} E_{i,j} \qquad (7)$$

N, is neighborhood number of a node. So if we consider the same network in Figure 1, we present by equation (8) the energy consumed by the four nodes in a period TSIM = 60 seconds with Hello Interval (AODV)( HI (AODV)) is equal to1 second according to the draft:

$$E_{total}(AODV) = \frac{4(nodes) * C * T_{SIM}}{HI(AODV)} \qquad (8)$$

In the case of PC-AODV and since almost the entire node performing the same transmission the resultant energy is given by:

$$E_{total}(PC-AODV) = E_R(A) + E_R(B) + E_R(C) + E_R(D)$$
$$= E_{HELLO} + 3 * E_{ACK} = \frac{4(nodes) * C * T_{SIM}}{HI(AODV)} \qquad (9)$$

Proof: let take HI (PC_AODV) = number of node responding with acknowledgment multiplied by HI(AODV). The node *A* that transmits a first hello message take the same interval as AODV protocol.

$$E_{HELLO} = \frac{C * T_{SIM}}{HI(AODV)}$$

$$E_{ACK} = \frac{3(nodes) * C * T_{SIM}}{HI(PC - AODV)} = \frac{3(nodes) * C * T_{SIM}}{3 * HI(AODV)} = \frac{C * T_{SIM}}{HI(AODV)}$$

According to equation (8) and (9) we note that the energy consumption due to the number of HELLO messages (adding the acknowledgment in the case of PC-AODV) is the same in both versions. So we do not degrade the classic version in terms of energy. Otherwise, our solution provides information about energy stored by the neighborhood under the same conditions.

## 7. SIMULATION RESULT

After the analytical comparison of the two protocols, we compare the number of messages exchanged using the simulators and the JIST / SWANS [13]. Simulations of these two protocols were made on the same simulations model (same parameters values and even traffic patterns).

The platform JIST/SWANS is a high-performance discrete event simulator, developed in Cornell University. JIST (Java in Simulation Time) is a driving simulation of discrete event which runs over a Java virtual machine (JVM). SWANS (Scalable Wireless Ad-hoc Network Simulator) are a network simulator that runs on top of JIST.

Through this simulator, we analyzed the overhead metric of the routing ad hoc protocol AODV and PC-AODV.

The Figure 3 shows the number of HELLO messages sent and received in case of AODV and on the same figure we also represented the sum of HELLO and HELLO_ACK messages sent and received in case of PC-AODV. Both of simulations are made in a period of 1800 seconds, with a variable number of nodes, static and randomly distributed nodes over an area of 1000x1000 meters. In Figure 4, 5 and 6 we have represented respectively Route Request, Route Reply and Route Error messages exchanged used by AODV and PC-AODV for route establishments. We also presented in figure 7 the routing overhead of AODV and PS-AODV.

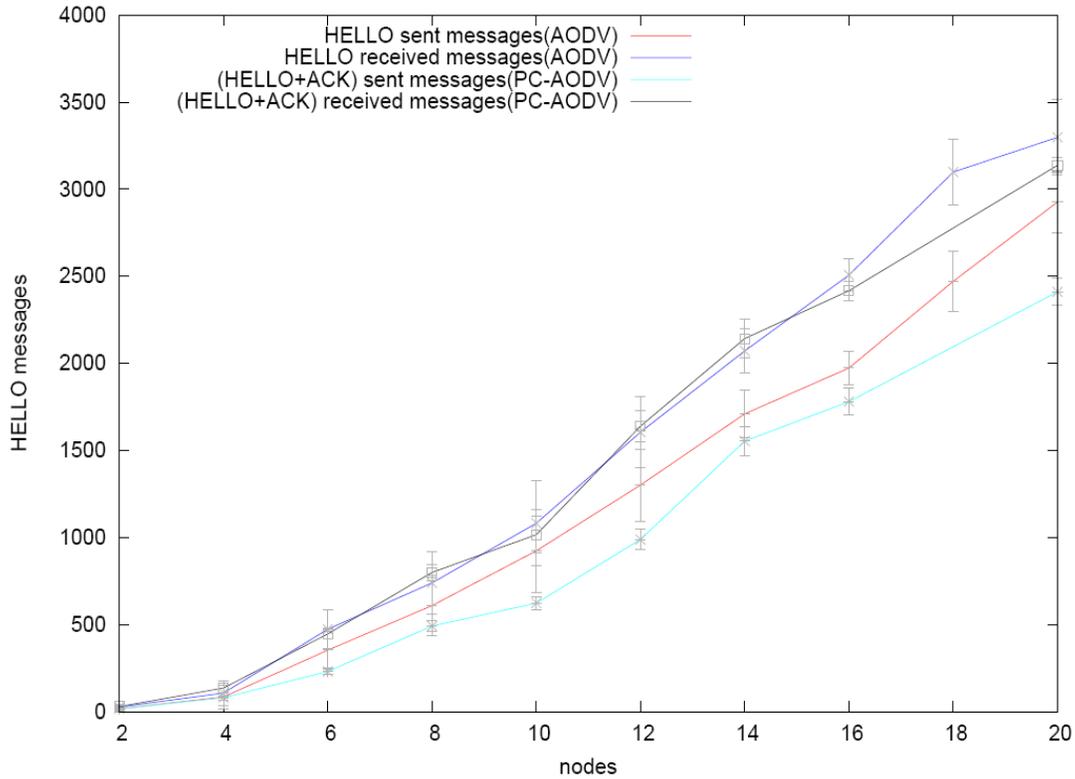

Figure 3. Exchanged Hello and Hello_ ACK messages

Through figure 3 we found that the number of HELLO messages in the case of AODV increases with the number of the node approximately linearly. Since the HELLO messages are periodic, so this increase depends of, the number of nodes, the period of HELLO messages, and the duration of simulation. Otherwise, the curve of HELLO messages reception is above than the curve of transmission and is solved by the fact that a hello message sent by one node will be received by the n neighbors of this node, and then a message is sent once and received more than once.

On the amounts of HELLO_ACK messages acknowledgment is slightly lower than the number of HELLO message, as we noted earlier, the nodes that has a very low energy cannot send acknowledgments in the period for reception of this type of messages. We used an energy model to assign a random power level for each node between two maximum and minimum thresholds, in order to have closely conditions of real network.

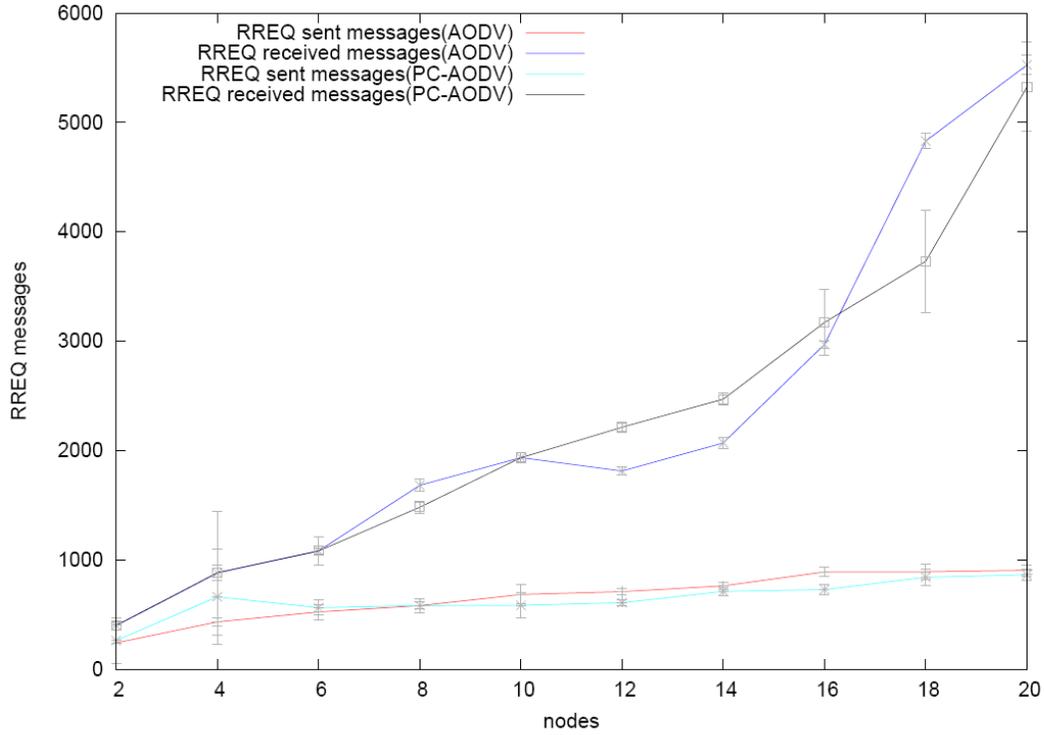

Figure 4. RouteREQuest Message exchanged

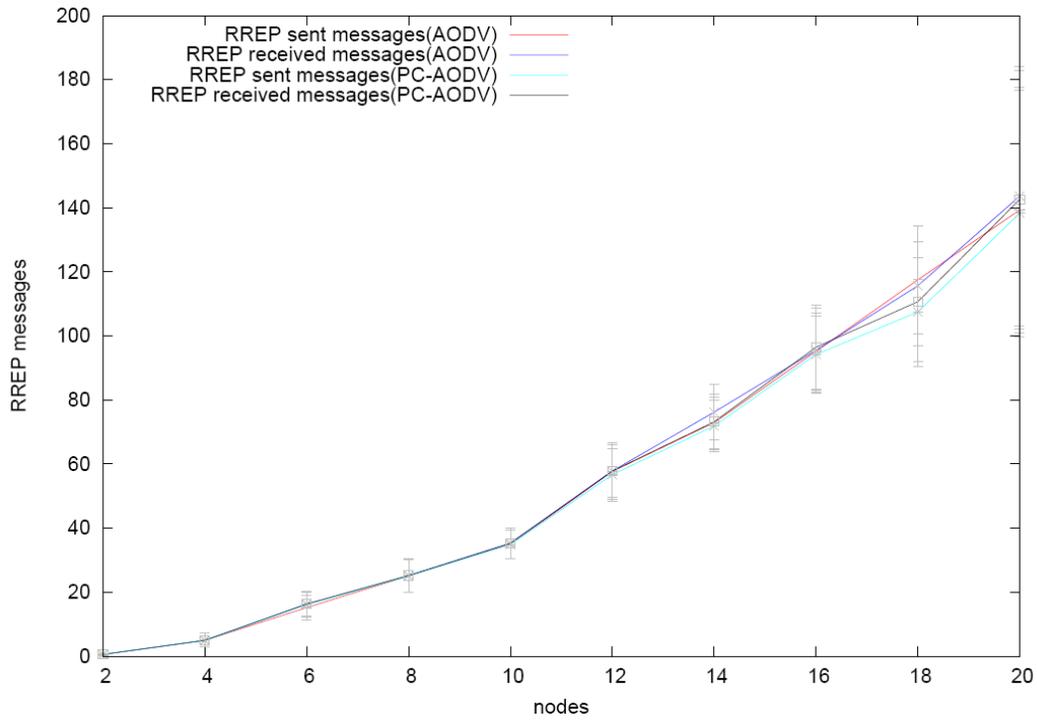

Figure 5. Exchanged Route REPlay messages

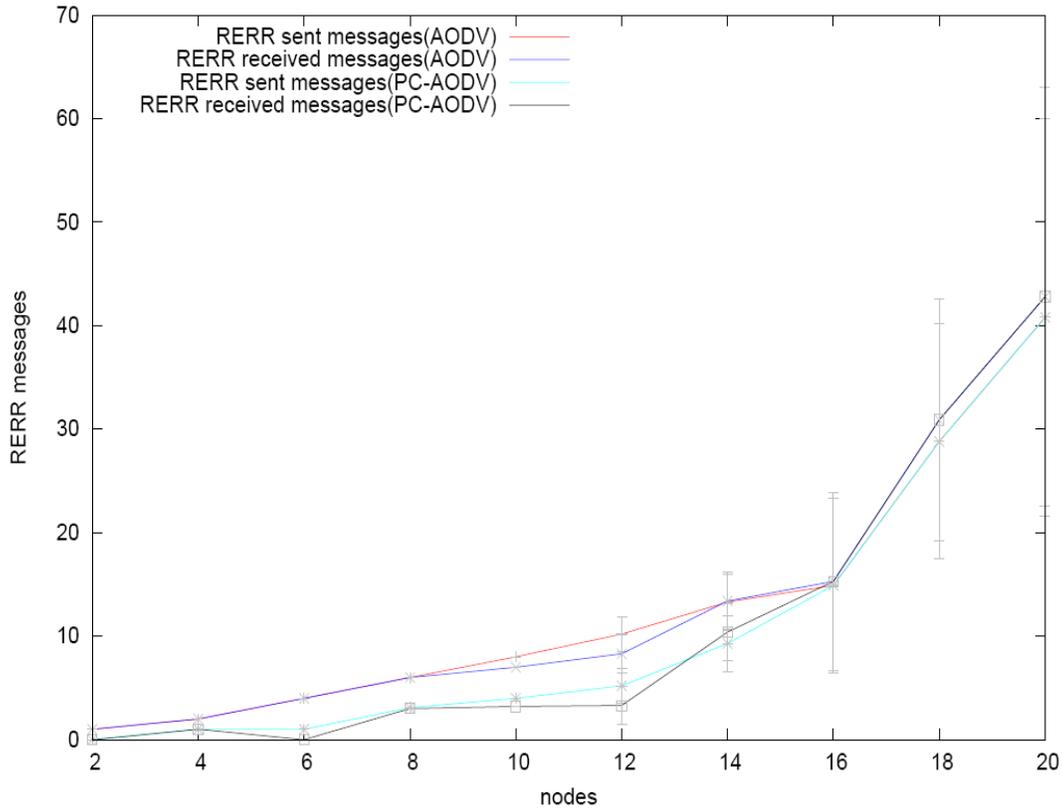

Figure 6. Exchanged Route ERRor messages

Figures 4, 5 and 6 show the influence of our solution on the number of control (other than Hello messages) exchanged. These two curves show that our solution allows AODV to generate the same traffic control and data but also shows that gains information about the energy stored in the batteries of the nodes.

So we note that the curves coincide with a slight difference due to the condition of random simulation to approximate a real network. The number of messages generated shows the overhead of network to establish paths and to send data packets. We can deduce two main metric, packet delivery ratio and overhead [31],[32],[33].

The letter is the most metric needed in Ad-Hoc network simulations. We found several definitions of this metric according to the parameters set of simulations. We chose one that measures the overhead of network control messages. It is equal to the number of control messages broadcast in the network divided by the sum of this number and the number of data messages sent. Overhead presented in Figure 7.

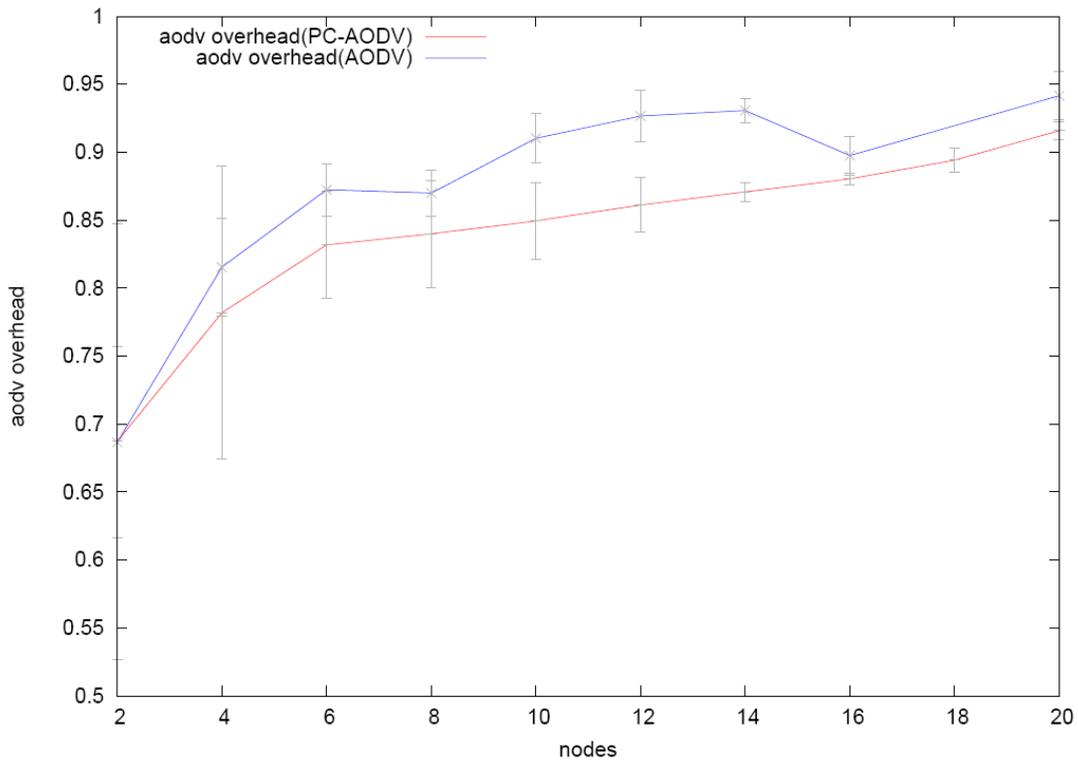

Figure 7. Routing overhead

Through this curve we first note that as the number of nodes increases, the network become charged, and it shows the disadvantage that AODV is not appropriate to the dense network. In addition the overhead of PC-AODV is lower than AODV, that because the number of HELLO message is smaller, so there was a slight decrease compared to AODV.

The packet delivery ratio is another mainly metric, it represent the safety of reception through the routing protocol used. It equal to number of received messages divided by the number of transmitted messages. We used the RREP messages, since this type of message follows the shortest path found (calculated using Dijekstra algorithm []) after the broadcast of RREQ messages. Packets delivery ratio is shown in Figure 8 using the equations (10):

$$PDR = \sum_{Transmission} \frac{\mathrm{Re}\,ceived\_Messages}{Transmitted\_Messages} \quad (10)$$

This solution provides to nodes to look more than shortest path, the safe path (the path that contains enough energy to routing packets). Indeed, using the batteries level information, the node doesn't choose a path that contain a node that risk to leave the network because it battery is empty. In the classical case, sometimes there was a link failure caused by the departure of a node due to the depletion of its battery, and this will reset again another route and that path we broadcast many more route discovery message.

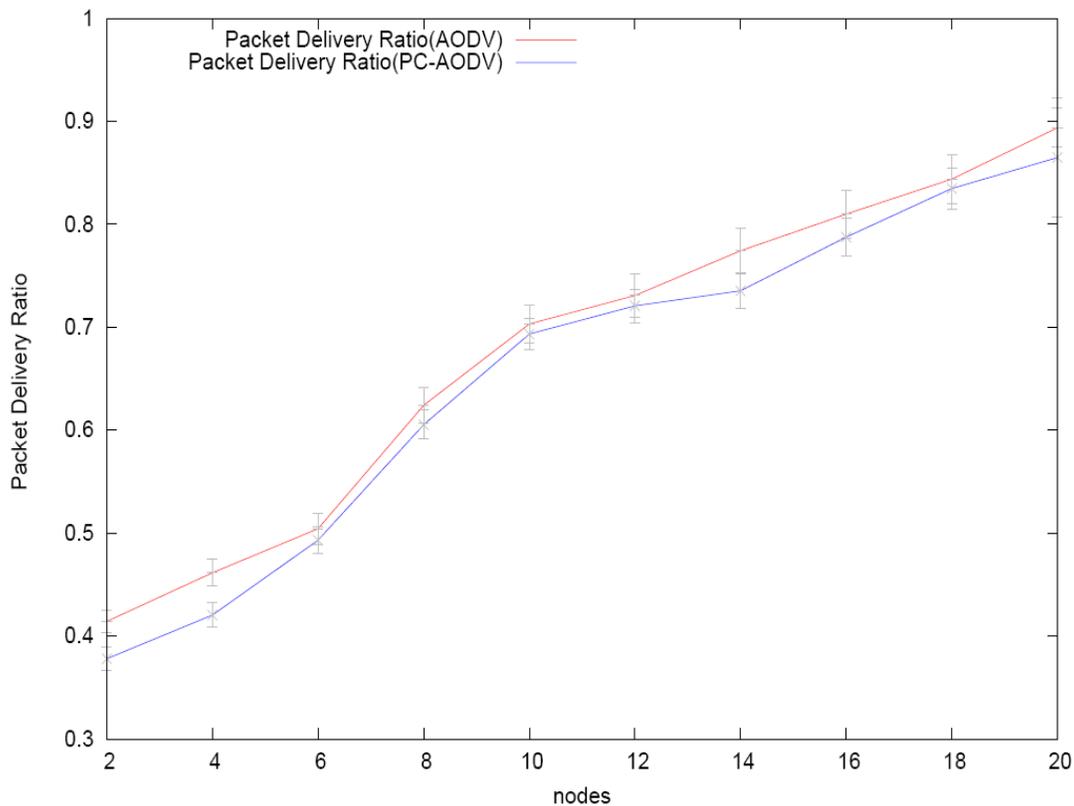

Figure 9. Packet Delivery Ratio

## 8. CONCLUSION:

In this paper we presented a new solution for the exchange of HELLO messages in AODV routing protocol. We have shown that our solution can provide knowledge about the levels of stored energy of the nodes constituting the network without affecting the operation of the protocol. After saving this new information, a node given in the network can choose the shortest path that contains enough energy for the correct routing of data packets, thus winning in terms of the ratio of the packets.

In future work, we will evaluate the PC-AODV performances in different topologies and different types of mobility to demonstrate the robustness of this Protocol and the benefits provided.

We will also set parameter that we guarantee the good receptions of acknowledgment, we will calculate the collision probability of acknowledgments to find factor to ensure the receipt of such messages, and therefore a hello sender node receives the acknowledgments messages in highly accurate moments.

**Authors**

**Dr. Maher HENI** was born in 1984 in KEF, Tunisia. He received the national engineering diploma in telecommunication from the National Engineering School of Tunis (ENIT) and the master diploma in Telecommunications from ENIT with collaboration of IEF institute in paris-sud university. Since 2010, he is a Ph.D. student in INNOV'COM (Innovation of COMmunicant and COoperative Mobiles Laboratory) in the Higher School of Communication (Sup'COM). His research interests are ad hoc networks and mobile communication systems. 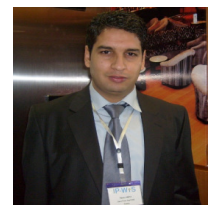

**Pr. Ridha BOUALLEGUE** was born in Tunis, Tunisia. He received the M.S degree in Telecommunications in 1990, the Ph.D. degree in telecommunications in 1994, and the HDR degree in Telecommunications in 2003, all from National School of engineering of Tunis (ENIT), Tunisia. Director and founder of National Engineering School of Sousse in 2005. Director of the School of Technology and Computer Science in 2010. Currently, Prof. Ridha Bouallegue is the director of Innovation of COMmunicant and COoperative Mobiles Laboratory,  INNOV'COM Sup'COM, Higher School of Communication. His current research interests include mobile and cooperative communications, Access technique, intelligent signal processing, CDMA, MIMO, OFDM and UWB systems. 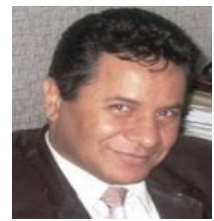